\documentclass[conference]{IEEEtran}
\IEEEoverridecommandlockouts
\usepackage{cite}
\usepackage{amsmath,amssymb,amsfonts}
\usepackage{algorithmic}
\usepackage{graphicx}
\usepackage{textcomp}
\usepackage{xcolor}
\usepackage{hyperref}
\usepackage{pgfplots}
\usepackage{subfigure}

\def\BibTeX{{\rm B\kern-.05em{\sc i\kern-.025em b}\kern-.08em
    T\kern-.1667em\lower.7ex\hbox{E}\kern-.125emX}}
    \pgfplotsset{compat=1.17}
\begin{document}

\title{Predicting and Visualizing Daily Mood of People Using Tracking Data of Consumer Devices and Services\\
}

\author{

\IEEEauthorblockN{Christian Reiser}
\IEEEauthorblockA{
\textit{University of Stuttgart}\\
Stuttgart, Germany \\
st141151@stud.uni-stuttgart.de\\
christian.reiser@insightme.org}





}

\maketitle

\begin{abstract}
Users can easily export personal data from devices (e.g., weather station and fitness tracker) and services (e.g., screentime tracker and commits on GitHub) they use but struggle to gain valuable insights. To tackle this problem, we present the self-tracking meta app called \textit{InsightMe}, which aims to show users how data relate to their wellbeing, health, and performance. This paper focuses on mood, which is closely associated with wellbeing. With data collected by one person, we show how a person's sleep, exercise, nutrition, weather, air quality, screentime, and work correlate to the average mood the person experiences during the day. Furthermore, the app predicts the mood via multiple linear regression and a neural network, achieving an explained variance of 55\% and 50\%, respectively. 
We strive for explainability and transparency by showing the users p-values of the correlations, drawing prediction intervals. In addition, we conducted a small A/B test on illustrating how the original data influence predictions. The source code\footnote{Sourcecode app: https://github.com/christianreiser/insightme\\
Sourcecode backend: https://github.com/christianreiser/correlate}
and app\footnote{Playstore: https://play.google.com/store/apps/details?id=com.insightme\\
Appstore: https://apps.apple.com/de/app/insightme/id1522480765}
are available online.
\end{abstract}


\section{Introduction}
We know that our environment and actions substantially affect our mood, health, intellectual and athletic performance.
However, there is less certainty about how much our environment (e.g., weather, air quality, noise) or behavior (e.g., nutrition, exercise, meditation, sleep) influence our happiness, productivity, sports performance, or allergies.
Furthermore, sometimes, we are surprised that we are less motivated, our athletic performance is poor, or disease symptoms are more severe.

This paper focuses on daily mood. Although negative moods have essential regulating functions like signaling the need for help or avoiding harmful behavior like going on buying sprees, taking risks, or making foolish investments \cite{noauthor_bipolar_nodate}, other studies show that bad moods can also have unfavorable consequences like less resistance to temptations, especially to unhealthy food\cite{fedorikhin_positive_2010}, impaired learning capabilities\cite{brand_how_2007}, and inhibited creative thinking\cite{vosburg_effects_1998}.

Our ultimate goal is to know which variables causally affect our mood to take beneficial actions. However, causal inference is generally a complex topic and not within the scope of this paper. Hence, we started with a system that computes how past behavioral and environmental data (e.g., weather, exercise, sleep, and screentime) correlate with mood and then use these features to predict the daily mood via multiple linear regression and a neural network. 
The system explains its predictions by visualizing its reasoning in two different ways. Version A is based on a regression triangle drawn onto a scatter plot, and version B is an abstraction of the former, where the slope, height, and width of the regression triangle are represented in a bar chart. We created a small A/B study to test which visualization method enables participants to interpret data faster and more accurately.



The data used in this paper come from inexpensive consumer devices and services which are passive and thus require minimal cost and effort to use.
The only manually tracked variable is the average mood at the end of each day, which was tracked via the app.


\section{Related Work}
This section provides an overview of relevant work, focusing on mood prediction (\ref{sec:wellbeing}) and related mobile applications with tracking, correlation, or prediction capabilities. (\ref{sec:apps}).

\subsection{Prediction of Mood}
\label{sec:wellbeing}

In the last decade, affective computing explored predicting mood, wellbeing, happiness, and emotion from sensor data gathered through various sources.

Harper and Southern\cite{harper_bayesian_2020} investigate how a unimodal heartbeat time series, measured with a professional EGC device, can predict emotional valence when the participant is seated. 

Choudhury et al. detect major depressive disorder of Twitter users who posted more than 4500 tweets on average with an average accuracy of \char`~ 70\%\cite{choudhury_predicting_2013}.

Several studies estimated mood, stress, and health with data from multimodal wearable sensors, a smartphone app, and daily manually reported behaviors such as academic activities and exercise, claiming maximum accuracies of
 68.48\%\cite{jaques_predicting_2015},
 74.3\%\cite{jaques_multi-task_2015},
 82.52\%\cite{jaques_multi-task_2016}, all with a baseline of 53.94. Another study scored 78.7\%, with a baseline of 50.4\%\cite{taylor_personalized_2020}.

All the studies mentioned above are less practical for non-professional users committed to long-term everyday usage because expensive professional equipment, time-consuming manual reporting of activity durations, or frequent social media behavior is needed.
Therefore, we focus on cheap and passive data sources, requiring minimal attention in everyday life.

One study meeting these criteria shows that mood can be predicted from passive data, specifically, keyboard and application data of mobile phones with a maximum accuracy of 66.59\% (62.65\% if without text)\cite{liu_multimodal_2020}. However, this project simplifies mood prediction to a classification problem with only three classes. Furthermore, compared to a high baseline of more than 43\% (due to class imbalance), the prediction accuracy of about 66\% is relatively low.\\

\subsection{Related Apps}
\label{sec:apps}

Several apps allow users to track their moods but lack correlation and prediction features  \cite{noauthor_moodprism_nodate}\cite{noauthor_moodily_nodate} \cite{noauthor_moodpanda_nodate}\cite{noauthor_daylio_nodate} \cite{noauthor_mood_nodate}\cite{noauthor_mood_nodate-1}\cite{appleinc_ios_nodate}\cite{hellocodeinc_exist_nodate}.
Some health apps allow correlating symptoms with food and behavior but still do not allow for prediction \cite{noauthor_pattern_nodate}\cite{noauthor_features_nodate}.

Apps capable of prediction are, and \cite{elitehrv_best_nodate} which estimates the activity of the sympathetic nervous system from heart rate and heart rate variability and \cite{welltory_welltory_nodate} which calculates stress, energy, and productivity levels from heart data as well\cite{welltory_welltory_nodate}.
Further, FitBit allows for logging how the user feels and computes a `Stress Management' score taking the manually logged feeling, data about sleep, electro-dermal activity, and exercise into account\cite{fitbit_stress_nodate}.
While these apps are capable of prediction, they are specialized in a few data types, which exclude mood, happiness, or wellbeing.

The product description of the smartwatch app `Happimeter' states to ``get your body signals to predict your mood with machine learning''\cite{noauthor_happimeter_nodate}. However, we could not test the app as it requires the operating system wearOS and the app has a user rating of only 1.5 of 5 stars on Google Play and was not updated for more than a year\cite{noauthor_happimeter_nodate}.


\section{Data sources}
This project aims to use non-intrusive, inexpensive sensors and services that are robust and easy to use for a few years.
Meeting these criteria, we tracked one person with a FitBit Sense smartwatch, indoor and outdoor weather stations, screentime logger, external variables like moon illumination, season, day of the week, manual tracking of mood, and more.
The reader can find a list of all data sources and explanations in the appendix (Section \ref{sec:Appendix}).

\section{Data Exploration and Processing}
This section describes how the data processing pipeline aggregates raw data, imputes missing data points, and exploits the past of the time series. Finally, we explore conspicuous patterns of some features.

\subsection{Pre-processing}
The sampling rates of the raw data typically vary between five minutes (e.g., heart rate) to about weekly (e.g., Bodyweight and $VO_2Max$).

\subsubsection{Data Aggregation}
The goal is to have a sampling rate of one sample per day.
In most cases, the sampling rate is greater than $1/24h$, and we aggregate the data to daily intervals by taking the sum, fifth percentile, 95th percentile, and median.
We use these percentiles instead of the minimum and maximum because they are less noisy and found them more predictive.

\subsubsection{Data Imputation}
The sampling frequency of Bodyweight and $VO_2Max$ is usually $<1/24h$.
Because Bodyweight and $VO_2Max$ represent physical entities that change relatively slowly, we assume a linear change, allowing linear interpolation of consecutive measurements to obtain the 24h frequency.
If there are days or features where many values are missing, we drop these days or features, respectively.
Otherwise, data imputation fills missing values with the feature's average.

\subsubsection{Time-series} 
\label{sec:time-series}
\begin{figure}[htbp]
\begin{center}
\includegraphics[width=1\linewidth]{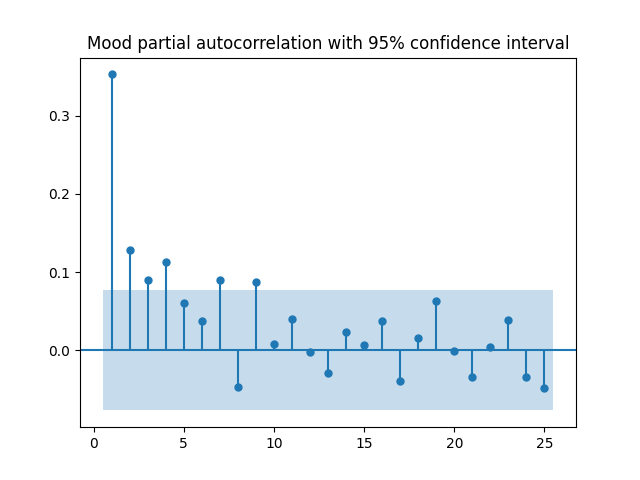}
\caption[]{Partial autocorrelation of mood. Values outside of the blue are within the 95\% confidence interval, thus statistically significant.}
\label{fig:partial_auto_corr}
\end{center}
\end{figure}

As the dataset is a time series, and yesterday's features could also affect today's mood, we added all of yesterday's features to the set of today's predictors.
We also include the mood of the last days until there is no new significant information about autocorrelation, given the mood of the previous days.
As shown in Figure \ref{fig:partial_auto_corr}, computing the partial autocorrelation\cite{durre_robust_2015} determines these days, when including all days from the left until the first insignificant day.
In our case, this means the values of one to four days ago.

\subsubsection{Standardization}
Standardization rescales the features to have a mean of 0 and unit variance.

\subsection{Data Exploration}
\begin{figure}[htbp]
\begin{center}
\includegraphics[width=1\linewidth]{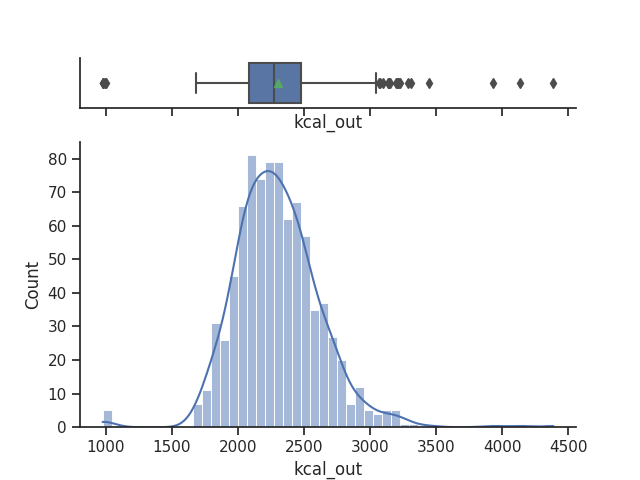}
\caption{The distribution of the estimated metabolic energy output. Values below 1600 kcal and above 3500 kcal are outliers.}
\label{fig:kcal_out}
\end{center}
\end{figure}

The dataset has many outliers because the sensors and services are cheap consumer devices.
For example, the estimated metabolic energy output, shown in Figure \ref{fig:kcal_out} has values at about 1000 kcal and above 4000 kcal.

\begin{figure}[htbp]
\begin{center}
\includegraphics[width=1\linewidth]{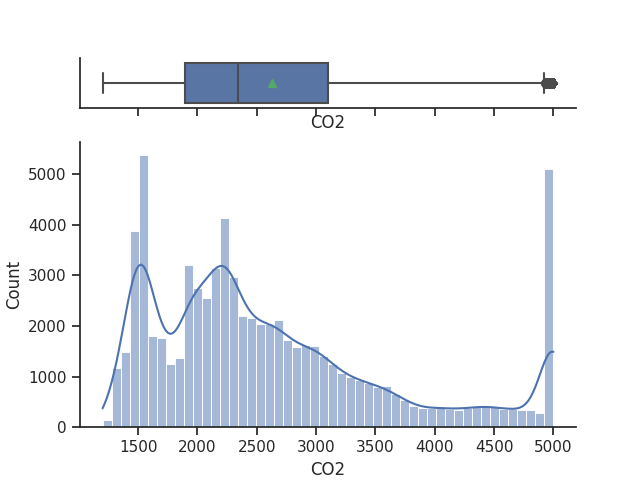}
\caption{Distribution of $CO_2$ data. Note the frequent occurrence of 5000ppm as it is the sensor's maximum range.}
\label{fig:co2}
\end{center}
\end{figure}

\begin{figure}[htbp]
\begin{tikzpicture}
\begin{axis}[
    title={CO2 Sensor data},
    xlabel={Timestamp},
    ylabel={CO2 in ppm},
    xmin=0, xmax=206,
    ymin=0, ymax=5500,
    xtick={0,50,100,150,200},
    ytick={0,1000,2000,3000,4000,5000},
    legend pos=south west,
    ymajorgrids=true,
    grid style=dashed,
]

\addplot[
    color=blue,
    mark=none,
    ]
    coordinates {
    (0,1782)(1,1778)(2,1751)(3,1753)(4,1737)(5,1737)(6,1752)(7,1697)(8,1736)(9,1715)(10,1723)(11,1792)(12,1962)(13,2057)(14,2175)(15,2250)(16,2355)(17,2415)(18,2560)(19,2603)(20,2704)(21,2734)(22,2835)(23,2883)(24,2977)(25,3085)(26,3185)(27,3253)(28,3321)(29,3393)(30,3441)(31,3502)(32,3572)(33,3651)(34,3275)(35,3267)(36,3284)(37,3345)(38,3434)(39,3486)(40,3568)(41,3606)(42,3688)(43,3792)(44,3831)(45,3902)(46,3923)(47,3996)(48,4031)(49,4075)(50,4153)(51,4171)(52,4245)(53,4325)(54,4366)(55,4366)(56,4401)(57,4406)(58,4447)(59,4464)(60,4469)(61,4511)(62,4566)(63,4630)(64,4649)(65,4649)(66,4727)(67,4745)(68,4838)(69,4815)(70,4877)(71,4914)(72,4936)(73,4936)(74,5000)(75,5000)(76,5000)(77,5000)(78,5000)(79,5000)(80,5000)(81,5000)(82,5000)(83,5000)(84,5000)(85,4979)(86,4918)(87,5000)(88,5000)(89,5000)(90,5000)(91,5000)(92,5000)(93,5000)(94,5000)(95,5000)(96,5000)(97,5000)(98,5000)(99,5000)(100,5000)(101,5000)(102,5000)(103,5000)(104,5000)(105,5000)(106,5000)(107,5000)(108,5000)(109,5000)(110,5000)(111,5000)(112,5000)(113,5000)(114,5000)(115,5000)(116,5000)(117,5000)(118,5000)(119,5000)(120,5000)(121,5000)(122,5000)(123,5000)(124,5000)(125,5000)(126,5000)(127,5000)(128,5000)(129,5000)(130,5000)(131,5000)(132,5000)(133,5000)(134,5000)(135,5000)(136,5000)(137,5000)(138,5000)(139,5000)(140,5000)(141,5000)(142,5000)(143,5000)(144,5000)(145,5000)(146,5000)(147,5000)(148,5000)(149,5000)(150,5000)(151,5000)(152,5000)(153,5000)(154,5000)(155,5000)(156,5000)(157,5000)(158,5000)(159,5000)(160,5000)(161,5000)(162,5000)(163,4250)(164,3647)(165,3256)(166,3034)(167,2885)(168,2687)(169,2629)(170,2600)(171,2653)(172,2657)(173,2637)(174,2707)(175,2773)(176,2870)(177,2918)(178,3017)(179,3068)(180,3087)(181,3203)(182,3238)(183,3288)(184,3348)(185,3383)(186,3104)(187,2819)(188,2646)(189,2586)(190,2625)(191,2757)(192,2862)(193,2888)(194,2969)(195,3039)(196,3051)(197,3140)(198,3175)(199,3211)(200,3175)(201,2933)(202,2729)(203,2597)(204,2541)(205,2492)(206,2393)
    };

\end{axis}
\end{tikzpicture}
    \caption{The actual CO2 level is higher than 5000 ppm, but the sensor's maximum is 5000 ppm.}
    \label{fig:netatmo_co2}
\end{figure}
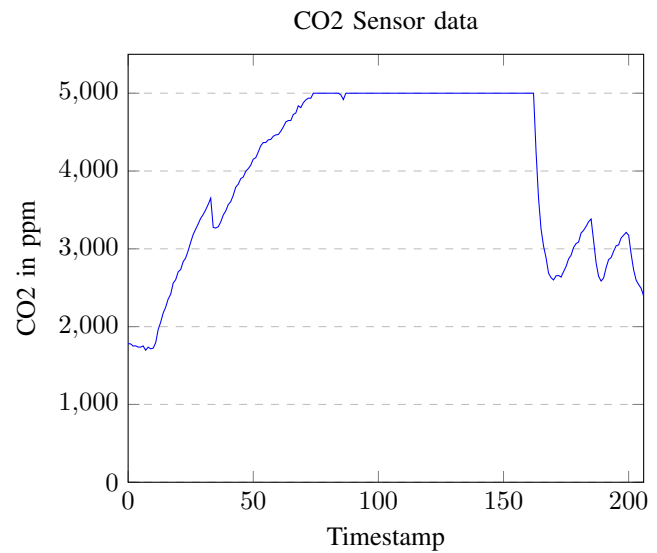

Moreover, Figure \ref{fig:co2} shows a suspicious $CO_2$ spike at 5000 ppm. A closer look into the raw sensor data depicted in Figure \ref{fig:netatmo_co2} indicates an improbable plateau at 5000 ppm. The causal explanation is an ending sensor range at 5000 ppm, which falsely counts all values greater than 5000 to 5000 ppm.

The distribution of the wakeup time looks gaussian except for one suspicious spike at 320 minutes after midnight. However, an alarm clock at 5:20 am indicates the plausibility of this spike.
\begin{figure}[htbp]
\begin{center}
\includegraphics[width=1\linewidth]{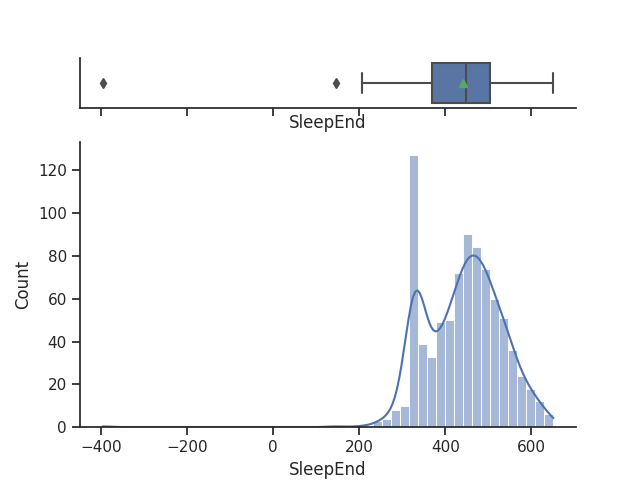}
\caption{Wakeup time in minutes after midnight. Note the spike at 320, which is caused by the alarm clock}.
\label{fig:alarm}
\end{center}
\end{figure}

Improbable values in the dataset are not corrected manually because we do not have access to data in the actual mobile application due to our strict privacy policy.
Instead, we exploit robust statistics by aggregating the data via the fifth and 95th percentile instead of the maxima. Our experiments have shown that these percentiles are more predictive than the maxima.

The app computes the Pearson correlation coefficient and p-values between all attributes. Because comparing each attribute with every other, we correct the p-values according to the Benjamini-Hochberg procedure\cite{benjamini_controlling_1995} to control the false discovery rate due to multiple testing.
We declare a result as significant for p $<$ 0.05.
Users can visually explore the data via a plotted time series with a seven-day moving average and manually inspect the relationship between two variables through scatter plots.

\subsection{Train-test-split}
Because there is a temporal dependency between observations, standard cross-validation, which assigns samples randomly to the train or test set, would lead to using some data from the future to forecast the past, which is not possible in the real-life application. 
We use time-series-splits to avoid this fallacy. However, many splits are inefficient w.r.t. data use because only the last split uses all training data. Ultimately, we create only one split, which is a simple train test split, where the training set contains old data points, and the test split the most recent ones.

\subsection{Multiple Linear Regression}
The estimation parameters for multiple linear regression are computed via the training dataset while applying elastic net regularization and sample-weighting.

\subsubsection{Regularization}
We use a combined $L_1$ and $L_2$ weight penalty called elastic net regularization\cite{zou_regularization_2005} with
\begin{equation}
\min _{w} \frac{1}{2 n_{\text {samples }}}\|X w-y\|_{2}^{2}+\alpha \rho\|w\|_{1}+\frac{\alpha(1-\rho)}{2}\|w\|_{2}^{2}
\end{equation}
being the objective function to minimize, where $\rho$ is the $L_1$ ratio, $\alpha$ the strength of the penalty terms, $w$ the weights, $X$ the features, and $y$ the predictions.
Specific cases are for $\rho=0$, which simplifies to ridge regularization and $\rho=1$ to lasso regularization. We search for the optimal $\rho$ and $\alpha$ via cross-validation.

\subsubsection{Sample-weighting}
\begin{figure}[htbp]
\begin{center}
\includegraphics[width=1\linewidth]{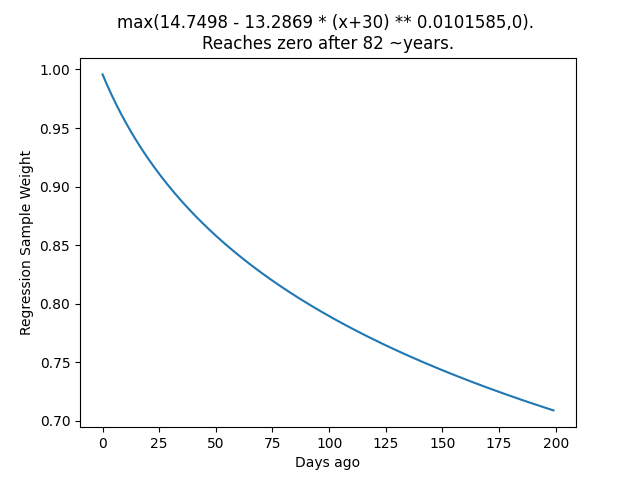}
\caption[Sample weight]{Exponential sample-weight decay discounting older data as they might be less valuable to predict today's mood.}
\label{fig:sample-weight}
\end{center}
\end{figure}

We assume that the factors influencing a person's mood change over time. To account for it, we use exponential sample-weight decay, as shown in Figure \ref{fig:sample-weight}. 
The formula is 
\begin{equation}
sample\_weight_i = max(14.7498-13.2869(i + 30)^{0.0101585},0),
\end{equation}
which is exponential decay fitted to a value of 1 for today's datapoint and reaches zero after about 82 years. $max(x,0)$ ensures that the sample weight never becomes a negative value. 

\subsection{Neural Network}
The neural network has two fully connected hidden layers, each with a leaky rectified linear unit\cite{xu_empirical_2015} as an activation function. The first and second hidden layers have 16 and 8 neurons, respectively. Because we want to regress on a scalar, the output layer has one unit. AdamW\cite{loshchilov_decoupled_2019} is the optimizer, minimizing the mean squared error for 4125 epochs with a learning rate of $10^{-4}$ and a weight decay of one. The number of epochs is determined by early stopping using cross-validation\cite{prechelt_automatic_1998}.
We searched for the best neural network architecture and hyper-parameters manually through cross-validation.



\section{Results}
Of 198 variables, 77 correlate significantly with mood. All correlation coefficients are in Table \ref{tab:features} in the appendix.

\subsection{Multiple Linear Regression}
Elastic net regularization with a penalty strength $\alpha=0.12$ and $L_1$-ratio $\rho=1$ leads to the best prediction performance on our dataset.

\begin{table}[]
\caption[]{Regression weights. All other features have a weight of zero.}
\begin{tabular}{ll}
\textbf{Feature}       & \textbf{Regression Weight} \\
HumidInMax()           & -0.116                     \\
HeartPoints            & \,\,0.101                      \\
CO2Median()            & -0.092                     \\
MoodYesterday          & \,\,0.065                      \\
NoiseMax()Yesterday    & -0.053                     \\
PressOutMin()Yesterday & \,\,0.050                      \\
BodyWeight             & \,\,0.046                      \\
VitaminDSup            & \,\,0.032                      \\
DistractingScreentime  & -0.009                    
\end{tabular}
\label{tab:reg}
\end{table}

Table \ref{tab:reg} shows all regression weights $w$ for $w\neq0$. Note that lasso regularization selected only nine features to predict the mood. 
The 95\% prediction interval is $\pm2.3$ on the original scale from 1 to 9 and $\pm1.2$ on the standardized scale with unit variance.
The average mean squared error on the standardized test set is 0.45, meaning 55\% of the original variance can be explained.
The effect of sample weighting is negligible.

\subsection{Neural Network}
The average mean squared error is 0.50 on the standardized test set, meaning half of the original variance can be explained.

\subsection{Explainability, Visualization, and A/B Test}
\label{sec:explain}
\begin{figure}[htbp]
    \centering
    \subfigure[Users can explore data via time-series with a seven-day moving average trend-line.]{
    \includegraphics[width=0.46\linewidth]{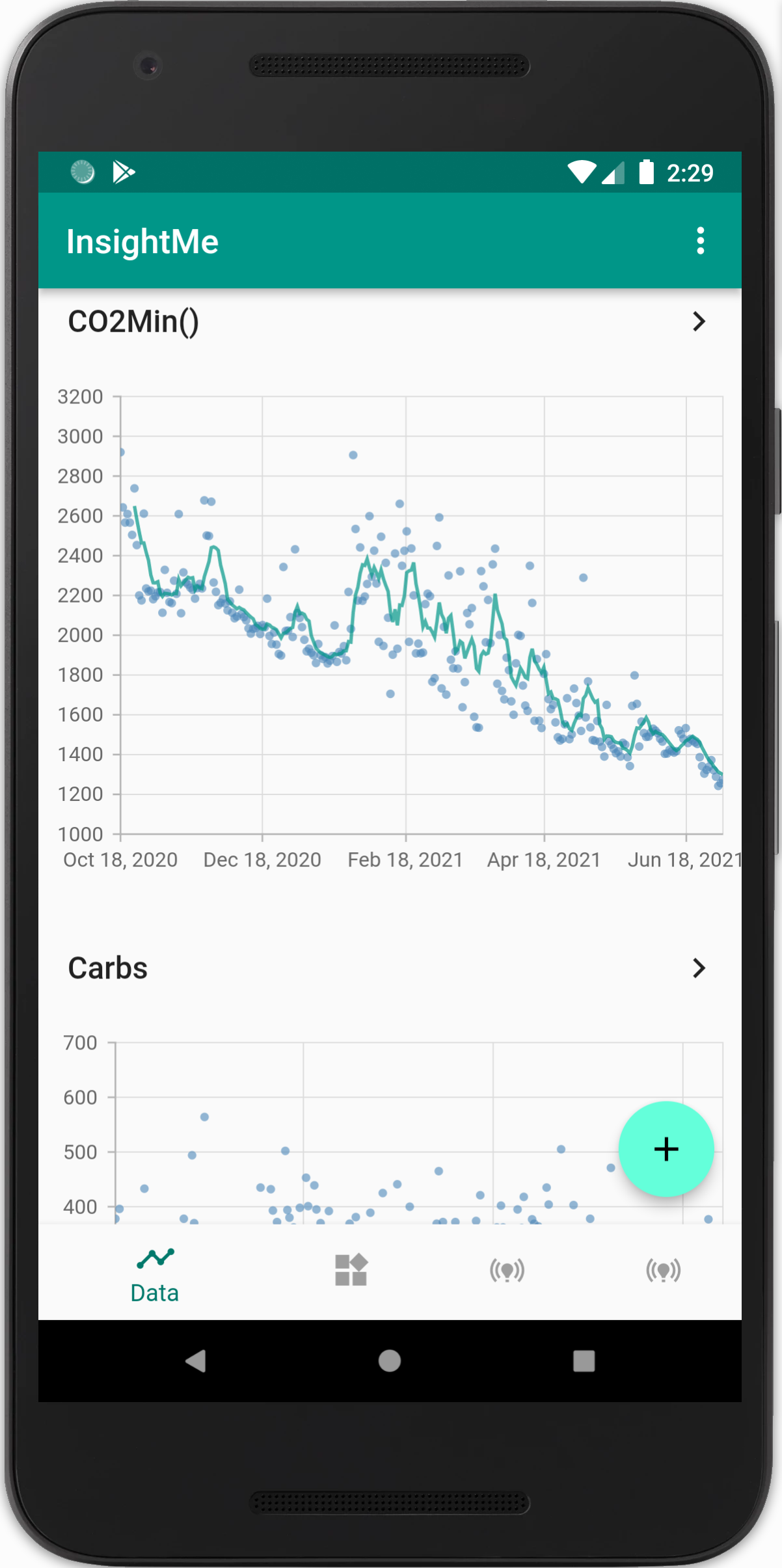}
    }
         \subfigure[A scatter plot with linear regression and Pearson correlation coefficient to visualize how variables relate to each other.]{
   \includegraphics[width=0.46\linewidth]{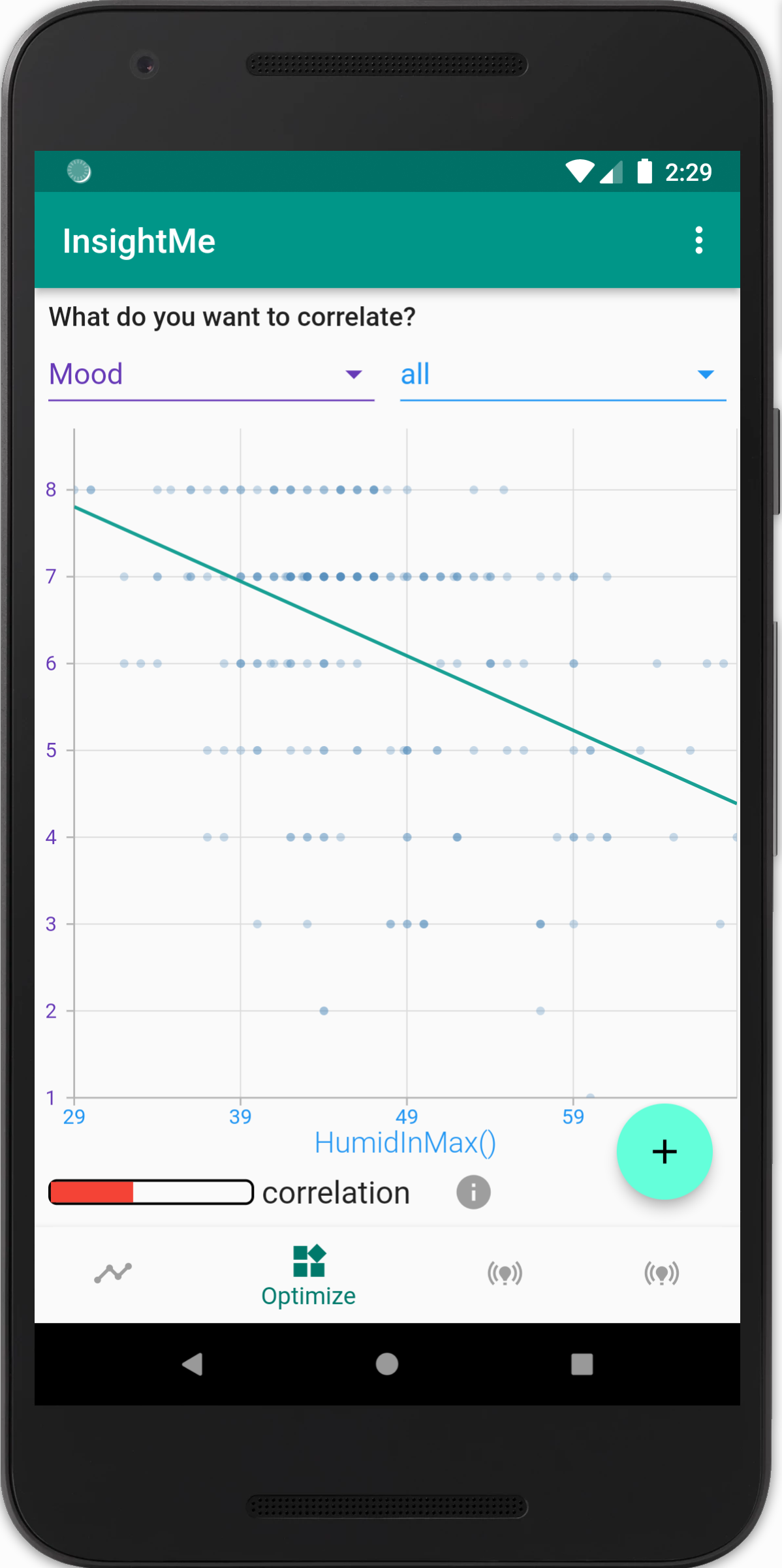}
}
\caption{Screenshots showing how users can explore data.}
    \label{fig:plot}
\end{figure}

Screenshot (a) of Figure \ref{fig:plot} shows how the app visualizes each feature via time series to allow the user to spot changes over time and trends through a seven-day moving average.
Screenshot (b) of Figure \ref{fig:plot} shows an example of a scatter plot enabling exploration of how variables visually relate to each other. In addition, it draws a linear regression line and indicates the degree of the linear relationship by visualizing the correlation coefficient in a bar.

\begin{figure}[htbp]
    \centering
    \includegraphics[width=0.6\linewidth]{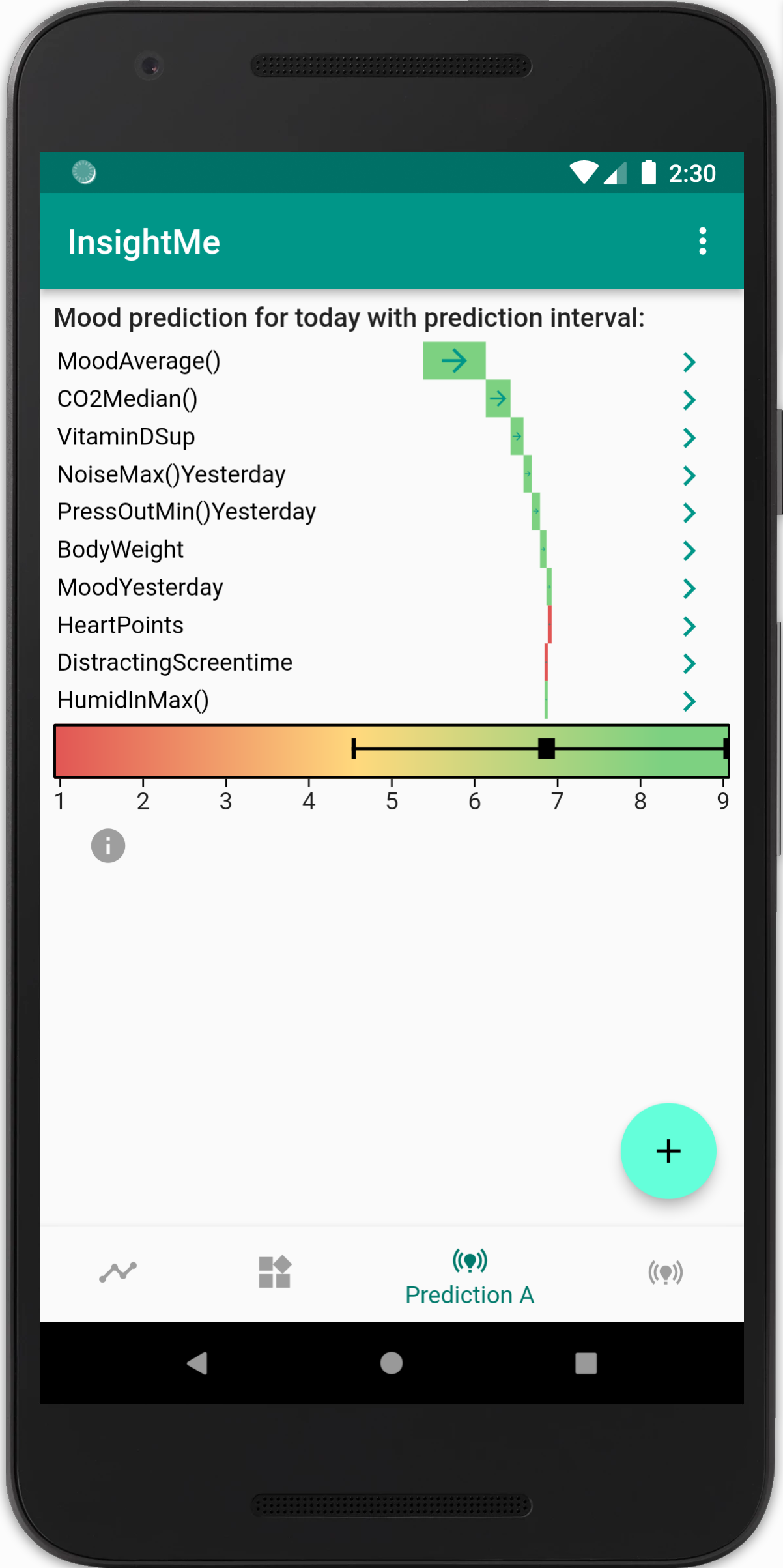}
    \caption{Mood prediction: Each row represents the contribution of a feature that was selected by $L_1$ regularization. The row contains a red or green bar if the contribution is negative or positive, and the size of the bar indicates the magnitude of the contribution.
    The black box with whiskers is the mood estimate with its 95\% prediction interval. 
    \label{fig:waterfall}
}
\end{figure}
Figure \ref{fig:waterfall} shows a black box with whiskers on a scale, which is the mood estimate with the 95\% prediction interval. This provides the users not only the prediction but also how much they can rely on the accuracy.
Above the prediction in screenshot\ref{fig:waterfall}, we explain to the user how multiple linear regression calculates the predictions. Each row represents the contribution of the selected features. The row contains a red or green bar if the contribution is negative or positive, respectively, and the size indicates the magnitude of the contribution. The final mood prediction is simply the sum of all contributions.

\begin{figure}[htbp]
    \centering
    \subfigure[]{
       \includegraphics[width=0.47\linewidth]{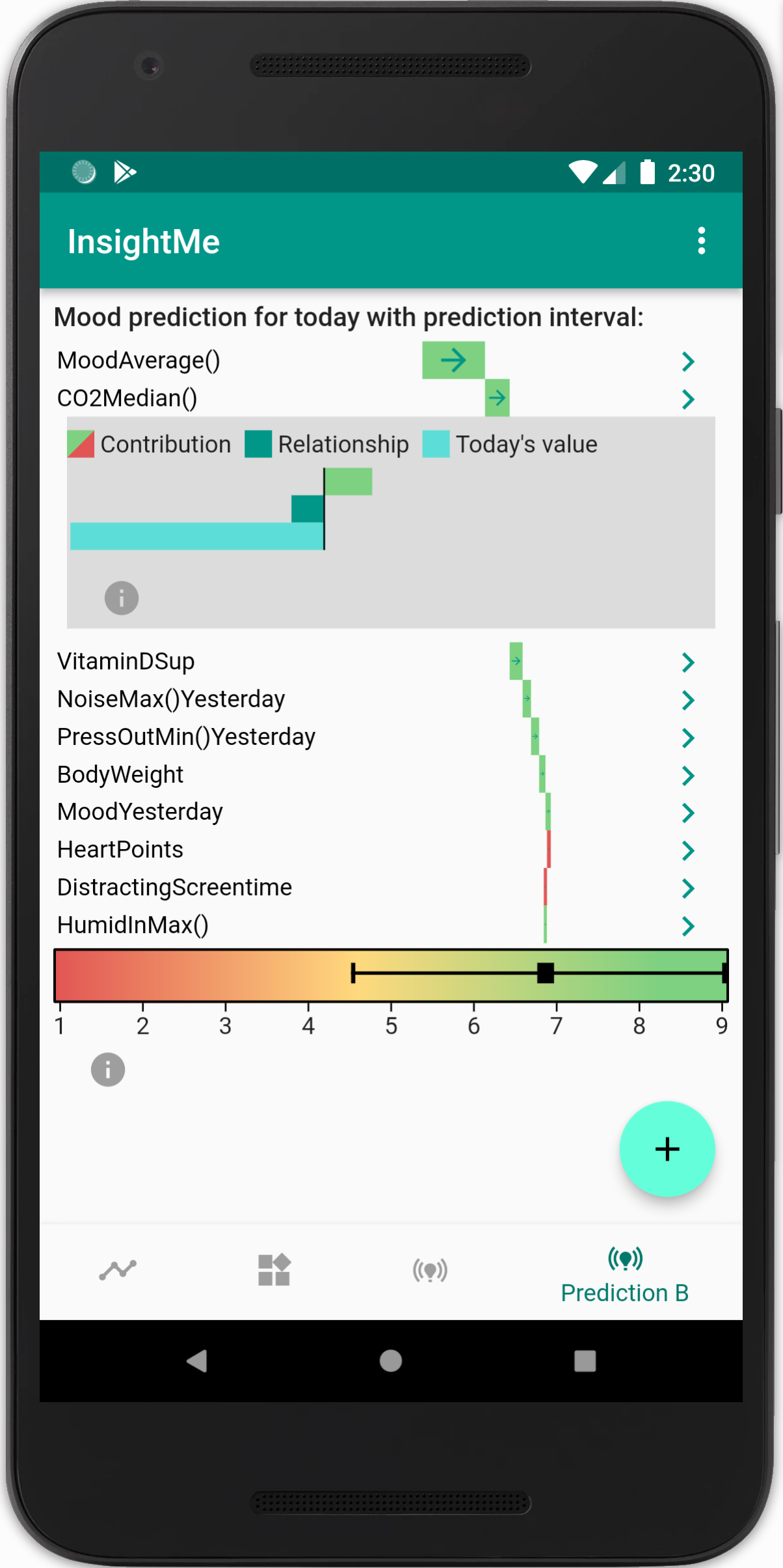}}
         \subfigure[]{
   \includegraphics[width=0.47\linewidth]{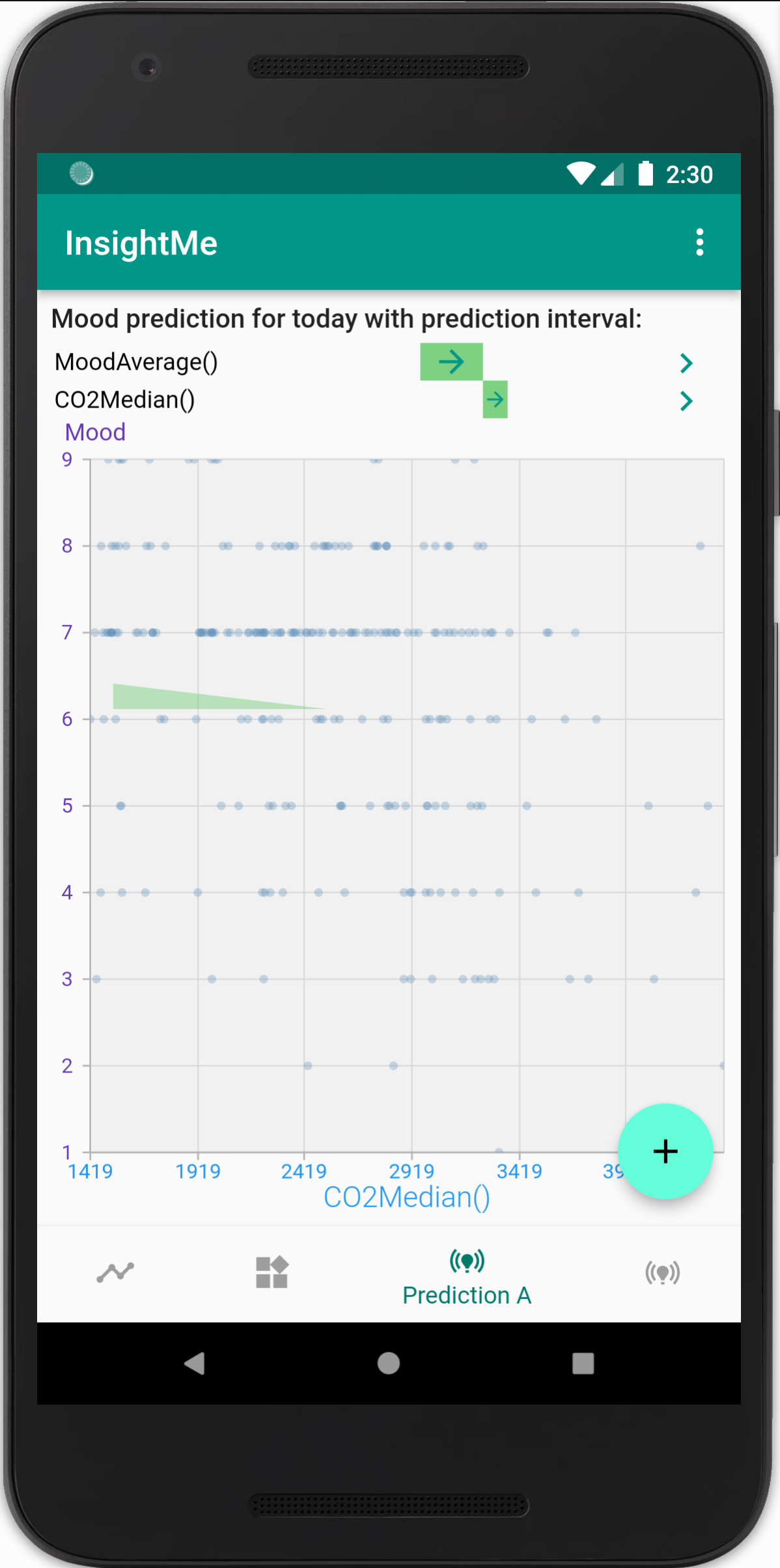}}
    
    \caption{Two methods explain how multiple linear regression computes the contribution of a single feature on the final prediction. We call the method (a) `bar chart' because it contains a chart with three bars and (b) `triangle chart', because it contains a red or green triangle inside a scatter plot.}
    \label{fig:explain}
\end{figure}

To understand more about the contribution of a feature, the user can tap on one and see either a bar chart or a regression triangle drawn onto the scatter plot.

The bar chart shows that the contribution (green or red bar) of a feature is the product of the weight of the feature (deep teal) and the difference from today's value to its average value (light teal). 

The triangle drawn onto the scatter plot shows the same information but with more context. 
The triangle's horizontal line represents the difference between the average and today's value of the feature. The triangle's slope represents the feature's weight, which is a regularized regression line of the two variables. Finally, the vertical length between both lines depicts the contribution.\\

The bar and triangle chart contain redundant information. Therefore, we conducted a small A/B test with 10 participants to determine which chart conveys the information more accurately and faster. 

Seven of the participants are male, and three are female. The age ranges from 19 to 32, and all of them have an engineering background. 
The participants answered four single choice questions concerning the bar chart (A) and four similar but not identical questions about the triangle chart (B). We assured proper testing by passively observing them. Each participant worked on both part A and part B; however, 50\% of the participants first completed part A and vice versa to control for the order. We measured the accuracy and time required to answer the questions of each part and reported their average w.r.t. the number of participants and number of questions. As shown in Table \ref{tab:study}, the `bar chart' results in a slightly higher accuracy of 90\% and 43 seconds a marginally faster completion, compared to the `regression triangle chart' with an accuracy of 85\% and 45 seconds.
While these results are not significant, the users also commented that they favor the `bar chart' as the length of the bar representing the weight is more accessible than the slope of the triangle, especially if the slope or triangle is small.

\begin{table}[]
\caption{Average required time and accuracy of single choice questions about the bar and triangle chart. The required time is relatively long because it includes the time to read the question.}
\begin{tabular}{lll}
\textbf{}             & bar chart & triangle chart \\
average accuracy      & 90\%       & 85\%           \\
average required time & 43 sec    & 45 sec        
\end{tabular}
    \label{tab:study}
\end{table}


\section{Discussion}

\subsubsection{Multicollinearity}
\begin{figure}[htbp]
\begin{center}
\includegraphics[width=1\linewidth]{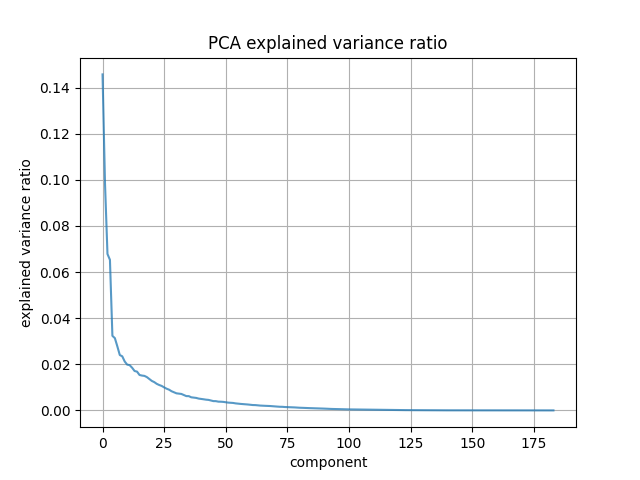}
\caption{The principal component analysis shows that there are only 25 components that explain more than 1\% of the variance, indicating highly dependent features.}
\label{fig:pca}
\end{center}
\end{figure}

Although 77 variables correlate significantly, there is multicollinearity. The principal component analysis shows in Figure \ref{fig:pca} that only 25 components explain more than 1\% of the total variance. Examples of correlated predictors are
\begin{itemize}
    \item all the fourfold aggregated variables (i.e., the mean, median, fifth-, and 95th percentile of a variable)
    \item weather indoor and outdoor
    \item time in bed and time asleep
    \item walking minutes, heart points, exertion points
\end{itemize}

\subsubsection{Multiple linear regression versus neural network}
Multiple linear regression performed better on the test set than the neural network. Neural networks can have an advantage over the linear multiple linear regression method because they can approximate nonlinear relationships. Still, the downside is the need for more training data to optimize additional parameters. In our case, the training set is probably too small, leading to overfitting and difficulty generalizing to new data. We expect the neural network will increase its performance with a growing dataset.
Besides, an enhanced architecture and improved hyperparameters could lead to better predictions of the neural network.
An advantage of multiple linear regression is good explainability, as illustrated in Figure\ref{fig:explain}, which is less intuitive for neural networks \cite{samek_explainable_2017}.

\subsection{Limitations}

\subsubsection{Large unexplained variance}
Predictions of the multiple linear regression and neural network leave about half the original variance unexplained. 
Factors limiting performance are:
\begin{itemize}
    \item unmeasured variables which influence a person's mood
    \item sensor data can be noisy, data imputation of missing values non-optimal
    \item multiple linear regression assumes a linear relationship between features and mood and neglects nonlinear mechanisms
    \item the training set might be too small, especially for the neural network
\end{itemize}

\subsubsection{Assumed linear mood scale}
Our method of asking the users' mood on an absolute scale from 1 to 9 assumes a linear relationship of these values. However, the genuine relationship might not be linear because there may be a higher degree of change on a comparative scale than on an absolute scale for extreme values\cite{carlsson_assessment_1983}.
Comparative surveys would reduce these biases. However, we decided against them because they require more of the user's time.

\subsubsection{Recency and fading affect bias}
Asking the user's average mood at the end of the day suffers from a potential recency bias\cite{cushing_barry_e_mitigation_1996}, where recent events of the evening have a stronger effect than more distant events in the morning. Furthermore, it is prone to the fading affect bias, where negative memories fade faster than positive ones\cite{skowronski_chapter_2014}.
Asking for a rating multiple times per day would reduce these biases; however, we decided against it because it requires more effort and is less sustainable over long periods.

\subsubsection{Survey}
The results of the survey are not conclusive because of the small number of participants. Furthermore, the selection of participants might not represent the actual distribution of the users, w.r.t. age, gender, and background.

\subsubsection{Causal inference}
Features correlating strongly with mood and predictors of multiple linear regression could potentially \textit{causally} affect mood. While this could be the case and interesting to know, correlation does not imply causation, and good predictors are not necessarily causes.
For example, the positive correlation between mood and exercise could be caused by exercise causing better mood, or that good mood leads someone to exercise more. Furthermore, both could affect each other, or there could be a confounder like good weather, which causes someone to exercise more and improve mood independently of each other.
In some instances, we can remove directions from a causal graph.
For example, when there is a positive correlation between the variable weekend and good mood, it is unlikely that the good mood causes the day to be a weekend. Unfortunately, this study does not allow for causal inference. Today, randomized controlled trials are still the gold standard for establishing causal conclusions\cite{hariton_randomised_2018}.

\subsection{Future Work}

\subsubsection{Stronger feature selection and weaker weight penalty}
\begin{figure}[htbp]
\begin{center}
\includegraphics[width=1\linewidth]{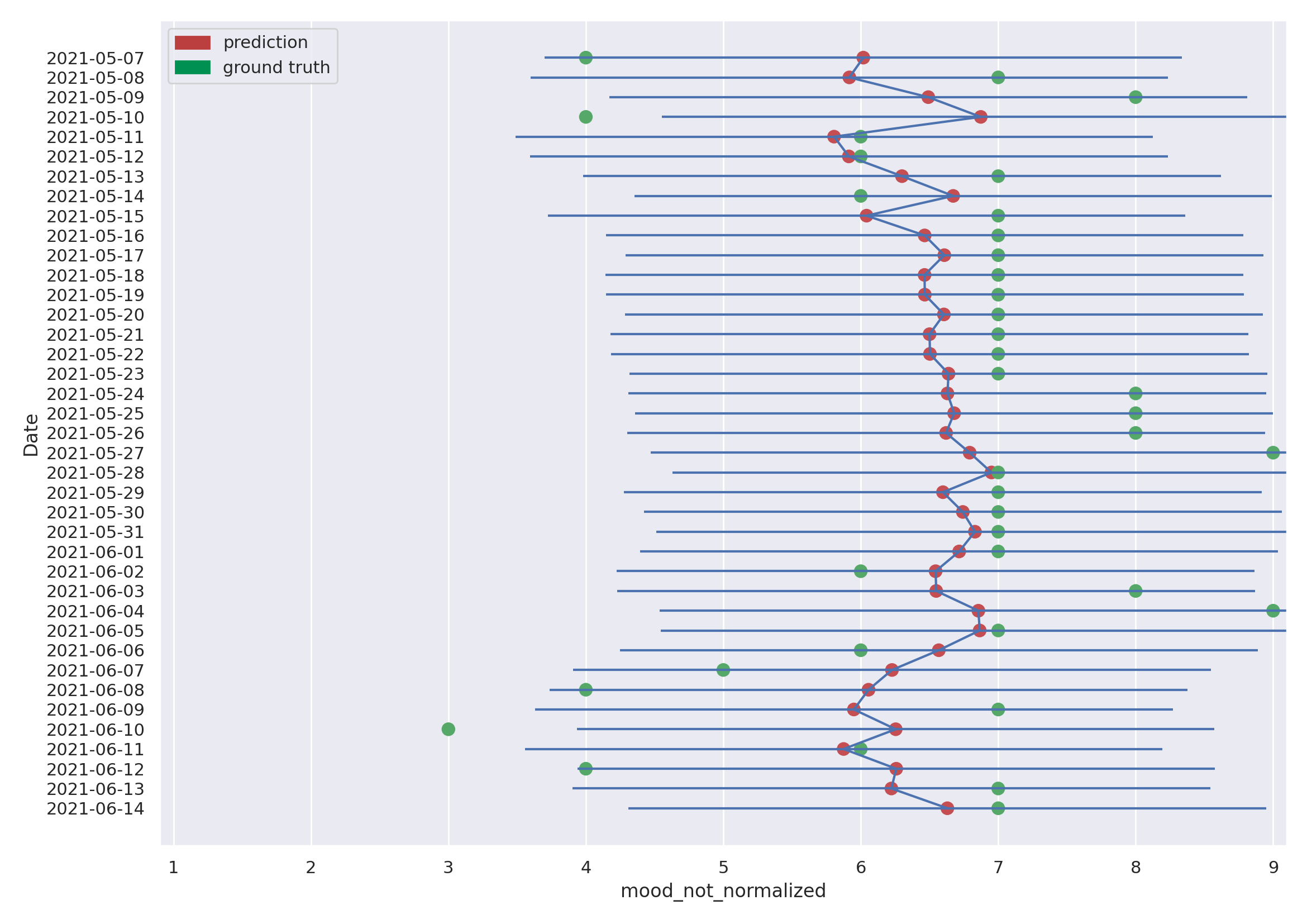}
\caption{Sample predictions: Predictions are in red, ground truth values in green. The blue bar represents the 95\% prediction interval. It seems that the predictions go in the right direction but are too close to the mean at 6.2.}
\label{fig:prediction-results}
\end{center}
\end{figure}
Figure \ref{fig:prediction-results}, shows sample predictions. It seems that predictions tend to go in the right direction but are too close to the mean at 6.2. We hypothesize that the weight penalty is too strong, which pushes predictions to the mean, but feature selection is critical because the overall regularization strength $\alpha \neq0$.
Further evidence is elastic's-net optimal $L_1$-ratio of $\rho=1$, meaning that performance is best with the lowest allowed weight penalty while maximizing feature selection. 

\begin{figure}[htbp]
\begin{center}
\includegraphics[width=1\linewidth]{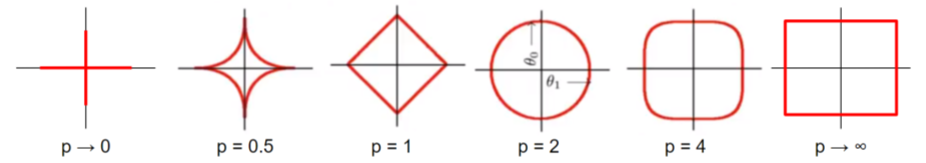}
\caption{Visualization of regularization with the $L_p$-norm. For $p\to 0$, the penalty is proportional to the number of non-zero parameters. For $p\to \infty$, the penalty has the size of the largest parameter. Elastic net regularization works only for $1 \leq p  \leq2$. Figure cropped and reprinted following the publisher's copyright\cite{binz_kevin_intro_2019}.}
\label{fig:reg}
\end{center}
\end{figure}

Fortunately, the equation of elastic net regularization, which operates between the $L_1$ and $L_2$-norm, can be generalized to the $L_p$-norm
\begin{equation}
\|x\|_{p}=\left(\sum_{i=1}^{n}\left|x_{i}\right|^{p}\right)^{1 / p},
\end{equation}
with $0<p<\infty$, allowing an even more extreme feature-selection to weight penalty ratio.
Figure \ref{fig:reg} shows how feature selection becomes stronger and the penalty for large parameters weaker as p gets closer to zero. Future work could explore if prediction performance improves for $p<1$ like applying a regularization with the $L_{0.5}$-norm. However, this has the downside of leaving convex optimization\cite{xu_l12_2012}.

\subsubsection{Improved data imputation}
 A common problem in long-term studies is that sensor data is missing. While we impute missing values with the average or linear interpolation, we plan to impute with a deep multimodal autoencoder to enable better mood prediction\cite{jaques_multimodal_2017}.

 \subsubsection{Forecasting tomorrow's mood}
This project explored predicting the mood of the same day, but we also plan to forecast tomorrow's mood. A study shows it is possible with a mean absolute error of 10.8 for workers and 17.8 for students, while the mood's standard deviation is 17.14\cite{umematsu_forecasting_2020}.

\subsection{Conclusion}
In this work, we presented a meta app that can import data from consumer devices and services and allows for manual tracking. The app allows the user to explore data via plotted time-lines and the relationship between variables through scatter plots and correlation coefficients.

The app predicts the mood or any other chosen target variable by automatically aggregating the data into daily features and selecting the best ones to predict the user's mood.
This project shows that multiple linear regression can explain more than half of the original variance.

We strive for transparency by conveying information about confidence through p-values and prediction intervals and created the app as an open-source project.

We hope the app helps users understand themselves better and improve their wellbeing, health, and physical \& cognitive performance.

\section{Acknowledgments}
I thank Benedikt V Ehinger and Tanja Blascheck for helpful discussions and suggestions.


\bibliographystyle{IEEEtran}
\bibliography{refs}


\section{Appendix}
\label{sec:Appendix}

Data sources are:
\begin{itemize}
\item Moon Illumination
\item daytime (the time between sunrise and sunset, which is shorter in the fall/winter and longer in the spring/summer)
\item OpenWeatherMap, which gives outside weather measurements GPS location 
   \begin{itemize}
\item temperature
\item heat index (human-perceived equivalent temperature, which includes humidity and windspeed)
\item air pressure
\item humidity
\item windspeed
\item cloud cover
\item precipitation (rain, snow)
\end{itemize}
\item Phone Location
   \begin{itemize}
\item longitude
\item latitude
\end{itemize}
\item Day of the week (Monday, Tuesday, etc.)
\item the severity of covid lockdown measures
\item commits on GitHub
\item Google Fit
   \begin{itemize}
\item Heart Points (duration and intensity exercise estimate)
\item walking minutes
\item running minutes
\item meditating minutes
\item Sleep efficiency
\item Sleep starting time
\item sleep end time
\item minutes in bed
\item minutes asleep
\item binary: was a nap taken?
\item minutes of light sleep
\item minutes of deep sleep
\item minutes of REM sleep
\item minutes awake in bed
\end{itemize}
\item Fitbit Charge 3 Smartwatch, which estimate:
   \begin{itemize}
\item steps taken
\item burned calories
\item heart rate
\item resting heart rate
\item VO2Max (maximum rate of oxygen consumption)
\item sleep revitalization
\item sleep duration score
\item sleep restlessness
\item vertical meters exercised (measured in floors)
\end{itemize}
\item Later Upgrade to Fitbit Sense Smartwatch, which extends the following estimates:
   \begin{itemize}
\item Responsiveness points: a proprietary assessment of how well the sympathetic and parasympathetic nervous system are in balance which takes Heart Rate Variability (HRV), Elevated Resting Heart Rate (RHR), Sleeping Heart Rate above RHR, and Electrodermal Activity (EDA) into account
\item Exertion points (similar to heart points)
\item the temperature of the wrist during sleep
\item sleep points (sleep rating)
\item stress points (average of sleep-, exertion-, and responsiveness points)
\end{itemize}
\item Netatmo indoor weather station
   \begin{itemize}
\item Temperature
\item Humidity
\item CO2 in ppm
\item Noise in dB
\item air pressure in Pa
\end{itemize}
\item Nutrition tracking with MyFitessPal
   \begin{itemize}
\item Carbohydrates intake
\item Fat intake
\item Protein intake
\item Sodium intake
\item fiber intake
\item sugar intake
\item cholesterol intake
\item total calories intake
\end{itemize}
\item Bodyweight (Renpho Scale) 
\item Screentime tracking, with RescueTime
   \begin{itemize}
\item total screen time
\item productive screentime
\item distracting screen time
\item neutral screen time
\end{itemize}
\end{itemize}

\begin{table}[]
\caption[]{Correlation coefficients, Benjamini–Hochberg corrected p-values and regression weights of all features. Sorted by p-values, which are bold when below 0.5.}
\label{tab:features}

\begin{tabular}{llll}
\textbf{feature} & \textbf{corr. coeff.} & \textbf{p-value} & \textbf{reg. weight} \\
MoodYesterday                  & 0.348  & \textbf{3.3E-18} & \textbf{0.06}  \\
HumidInMax()                   & -0.406 & \textbf{4.6E-09} & \textbf{-0.12} \\
BodyWeight                     & 0.248  & \textbf{4.6E-09} & \textbf{0.05}  \\
MoodEreyesterday               & 0.236  & \textbf{2.9E-08} & 0.00           \\
BodyWeightYesterday            & 0.235  & \textbf{3.0E-08} & 0.00           \\
Covid\_lifestyleYesterday      & 0.211  & \textbf{1.1E-06} & 0.00           \\
Covid\_lifestyle               & 0.210  & \textbf{1.1E-06} & 0.00           \\
HumidInMedian()                & -0.342 & \textbf{1.4E-06} & 0.00           \\
HumidInMax()Yesterday          & -0.329 & \textbf{4.2E-06} & 0.00           \\
CO2Median()                    & -0.321 & \textbf{7.3E-06} & \textbf{-0.09} \\
HumidOutMin()Yesterday         & -0.186 & \textbf{2.2E-05} & 0.00           \\
NoiseMax()Yesterday            & -0.303 & \textbf{2.9E-05} & \textbf{-0.05} \\
TempInMedian()                 & -0.301 & \textbf{3.0E-05} & 0.00           \\
HumidOutMean()Yesterday        & -0.181 & \textbf{3.3E-05} & 0.00           \\
HumidOutMax()Yesterday         & -0.179 & \textbf{3.8E-05} & 0.00           \\
HumidOutMin()                  & -0.177 & \textbf{5.1E-05} & 0.00           \\
NoiseMax()                     & -0.288 & \textbf{6.8E-05} & 0.00           \\
Daytime                        & 0.172  & \textbf{8.1E-05} & 0.00           \\
DaytimeYesterday               & 0.171  & \textbf{8.1E-05} & 0.00           \\
TempOutMax()Yesterday          & 0.171  & \textbf{8.7E-05} & 0.00           \\
HumidOutMean()                 & -0.168 & \textbf{1.2E-04} & 0.00           \\
HumidInMedian()Yesterday       & -0.276 & \textbf{1.4E-04} & 0.00           \\
TempInMax()                    & -0.270 & \textbf{2.0E-04} & 0.00           \\
HumidOutMax()                  & -0.162 & \textbf{2.0E-04} & 0.00           \\
TempOutMean()Yesterday         & 0.160  & \textbf{2.4E-04} & 0.00           \\
TempOutMax()                   & 0.157  & \textbf{3.2E-04} & 0.00           \\
HRMean()Yesterday              & -0.155 & \textbf{3.8E-04} & 0.00           \\
TempInMax()Yesterday           & -0.258 & \textbf{3.8E-04} & 0.00           \\
HumidInMin()                   & -0.257 & \textbf{3.8E-04} & 0.00           \\
TempInMedian()Yesterday        & -0.250 & \textbf{6.2E-04} & 0.00           \\
TempOutMean()                  & 0.148  & \textbf{7.8E-04} & 0.00           \\
TempFeelOutMax()Yesterday      & 0.147  & \textbf{7.8E-04} & 0.00           \\
TempFeelOutMax()               & 0.145  & \textbf{9.6E-04} & 0.00           \\
TempInMin()                    & -0.240 & \textbf{1.0E-03} & 0.00           \\
VitaminDSupYesterday           & 0.144  & \textbf{1.0E-03} & 0.00           \\
VitaminDSup                    & 0.143  & \textbf{1.1E-03} & \textbf{0.03}  \\
TempOutFeelMean()Yesterday     & 0.142  & \textbf{1.1E-03} & 0.00           \\
PressureInMin()Yesterday       & 0.234  & \textbf{1.3E-03} & 0.00           \\
TempOutFeelMean()              & 0.134  & \textbf{2.5E-03} & 0.00           \\
TempOutDelta()Yesterday        & 0.133  & \textbf{2.6E-03} & 0.00           \\
CO2Max()                       & -0.220 & \textbf{2.8E-03} & 0.00           \\
PressureInMedian()Yesterday    & 0.221  & \textbf{2.8E-03} & 0.00           \\
Fiber                          & 0.195  & \textbf{3.6E-03} &                \\
TempFeelOutMin()Yesterday      & 0.129  & \textbf{3.8E-03} & 0.00           \\
TempOutMin()Yesterday          & 0.127  & \textbf{4.5E-03} & 0.00           \\
PressureInMax()Yesterday       & 0.209  & \textbf{4.9E-03} & 0.00           \\
FiberYesterday                 & 0.188  & \textbf{5.3E-03} &                \\
TempOutDelta()                 & 0.124  & \textbf{5.3E-03} & 0.00           \\
SleepRevitalizationYesterday   & 0.123  & \textbf{6.3E-03} & 0.00           \\
HeartPoints                    & 0.121  & \textbf{6.6E-03} & \textbf{0.10}  \\
HighLongitude                  & 0.120  & \textbf{6.6E-03} & 0.00           \\
CO2Min()                       & -0.199 & \textbf{7.0E-03} & 0.00           \\
CO2Median()Yesterday           & -0.199 & \textbf{7.2E-03} & 0.00           \\
TempFeelOutMin()               & 0.117  & \textbf{8.6E-03} & 0.00           \\
SleepLight                     & -0.117 & \textbf{8.6E-03} & 0.00           \\
HumidInMin()Yesterday          & -0.194 & \textbf{8.9E-03} & 0.00           \\
TempOutMin()                   & 0.115  & \textbf{9.2E-03} & 0.00           \\
CloudOutMean()                 & -0.114 & \textbf{1.0E-02} & 0.00           \\
CloudOutMean()Yesterday        & -0.114 & \textbf{1.0E-02} & 0.00           \\
CloudOutMax()Yesterday         & -0.113 & \textbf{1.1E-02} & 0.00           \\
NoiseMedian()Yesterday         & -0.188 & \textbf{1.1E-02} & 0.00           \\
TempInMin()Yesterday           & -0.185 & \textbf{1.3E-02} & 0.00           \\
NoiseMedian()                  & -0.182 & \textbf{1.4E-02} & 0.00           \\
StressManagement               & 0.179  & \textbf{1.5E-02} & 0.00           \\
CO2Max()Yesterday              & -0.181 & \textbf{1.5E-02} & 0.00           \\
DistractingScreentimeYesterday & 0.110  & \textbf{1.5E-02} & 0.00           \\
PressureInMin()                & 0.174  & \textbf{2.0E-02} & 0.00           \\
LowLongitude                   & 0.103  & \textbf{2.2E-02} & 0.00           \\
ExertionPointsYesterday        & 0.169  & \textbf{2.2E-02} & 0.00           \\
PressureInMedian()             & 0.165  & \textbf{2.9E-02} & 0.00           \\
CarbsYesterday                 & 0.149  & \textbf{2.9E-02} &                \\
\end{tabular}
\end{table}
\begin{table}[]
\begin{tabular}{llll}
\textbf{feature} & \textbf{corr. coeff.} & \textbf{p-value} & \textbf{reg. weight} \\
CO2Min()Yesterday              & -0.164 & \textbf{3.0E-02} & 0.00           \\
SleepDeep                      & 0.094  & \textbf{3.8E-02} & 0.00           \\
PressureInMax()                & 0.153  & \textbf{4.6E-02} & 0.00           \\
HighLongitudeYesterday         & 0.089  & 5.4E-02          & 0.00           \\
SleepStartYesterday            & -0.089 & 5.4E-02          & 0.00           \\
HRResting                      & -0.085 & 6.5E-02          & 0.00           \\
CloudOutMax()                  & -0.084 & 7.2E-02          & 0.00           \\
RunningMin                     & 0.082  & 7.8E-02          & 0.00           \\
LowLongitudeYesterday          & 0.082  & 7.8E-02          & 0.00           \\
SleepPoints                    & 0.134  & 8.3E-02          & 0.00           \\
ScreentimeYesterday            & 0.083  & 8.3E-02          & 0.00           \\
StressManagementYesterday      & 0.131  & 9.0E-02          & 0.00           \\
NeutralScreentimeYesterday     & 0.080  & 9.5E-02          & 0.00           \\
SodiumYesterday                & 0.118  & 9.6E-02          &                \\
SleepRevitalization            & 0.077  & 1.1E-01          & 0.00           \\
Sodium                         & 0.113  & 1.2E-01          &                \\
HRMax()                        & 0.072  & 1.3E-01          & 0.00           \\
SleepLightYesterday            & -0.072 & 1.4E-01          & 0.00           \\
LowLatitude                    & -0.071 & 1.4E-01          & 0.00           \\
RunningMinYesterday            & 0.071  & 1.4E-01          & 0.00           \\
ExertionPoints                 & 0.108  & 1.8E-01          & 0.00           \\
SugarYesterday                 & 0.099  & 1.8E-01          &                \\
HRMax()Yesterday               & 0.065  & 1.9E-01          & 0.00           \\
CloudOutMin()                  & -0.065 & 1.9E-01          & 0.00           \\
Steps                          & 0.060  & 2.4E-01          & 0.00           \\
SleepWake                      & -0.060 & 2.4E-01          & 0.00           \\
SleepEndYesterday              & 0.059  & 2.5E-01          & 0.00           \\
SleepREM                       & 0.056  & 2.8E-01          & 0.00           \\
MeditatingMinYesterday         & 0.056  & 2.9E-01          & 0.00           \\
Floors                         & 0.056  & 2.9E-01          & 0.00           \\
KCalIn                         & 0.083  & 2.9E-01          &                \\
HRMin()Yesterday               & -0.054 & 3.0E-01          & 0.00           \\
SleepREMYesterday              & 0.052  & 3.3E-01          & 0.00           \\
RainSnowYesterday              & -0.051 & 3.4E-01          & 0.00           \\
PressOutDelta()Yesterday       & -0.051 & 3.4E-01          & 0.00           \\
MeditatingMin                  & 0.050  & 3.5E-01          & 0.00           \\
InBedMin                       & -0.050 & 3.5E-01          & 0.00           \\
HRRestingYesterday             & -0.050 & 3.5E-01          & 0.00           \\
HRMin()                        & -0.048 & 3.8E-01          & 0.00           \\
Sunday                         & 0.047  & 3.9E-01          & 0.00           \\
SaturdayYesterday              & 0.047  & 3.9E-01          & 0.00           \\
Carbs                          & 0.069  & 4.0E-01          &                \\
Saturday                       & -0.044 & 4.2E-01          & 0.00           \\
FridayYesterday                & -0.044 & 4.2E-01          & 0.00           \\
SleepStart                     & -0.043 & 4.3E-01          & 0.00           \\
SleepEfficiencyYesterday       & -0.043 & 4.3E-01          & 0.00           \\
SleepWakeYesterday             & -0.043 & 4.3E-01          & 0.00           \\
NoiseMin()Yesterday            & 0.071  & 4.4E-01          & 0.00           \\
PressOutDelta()                & -0.042 & 4.4E-01          & 0.00           \\
CloudOutMin()Yesterday         & -0.041 & 4.5E-01          & 0.00           \\
HeartPointsYesterday           & 0.041  & 4.5E-01          & 0.00           \\
PressOutMin()Yesterday         & 0.041  & 4.5E-01          & \textbf{0.05}  \\
HighLatitude                   & -0.041 & 4.5E-01          & 0.00           \\
NoiseMin()                     & 0.068  & 4.5E-01          & 0.00           \\
ProductiveScreentimeYesterday  & 0.041  & 4.6E-01          & 0.00           \\
ProteinYesterday               & 0.060  & 4.6E-01          &                \\
ResponsivenessPoints           & 0.065  & 4.6E-01          & 0.00           \\
LowLatitudeYesterday           & -0.039 & 4.6E-01          & 0.00           \\
WindOutMean()Yesterday         & 0.039  & 4.6E-01          & 0.00           \\
InBedMinYesterday              & -0.039 & 4.6E-01          & 0.00           \\
FatYesterday                   & -0.058 & 4.7E-01          &                \\
KcalOut                        & 0.038  & 4.7E-01          & 0.00           \\
Monday                         & -0.037 & 4.7E-01          & 0.00           \\
SundayYesterday                & -0.037 & 4.7E-01          & 0.00           \\
WindOutMin()Yesterday          & -0.037 & 4.7E-01          & 0.00           \\
HRMean()                       & -0.037 & 4.7E-01          & 0.00           \\
Cholesterol                    & 0.056  & 4.7E-01          &                \\
AsleepMin                      & -0.035 & 5.0E-01          & 0.00           \\
NeutralScreentime              & 0.035  & 5.1E-01          & 0.00           \\
KCalInYesterday                & 0.052  & 5.2E-01          &                \\
Sugar                          & 0.051  & 5.2E-01          &                \\
DistractingScreentime          & 0.034  & 5.2E-01          & \textbf{-0.01} \\
VO2MaxYesterday                & 0.034  & 5.2E-01          & 0.00           \\
RainSnow                       & -0.033 & 5.2E-01          & 0.00           \\
Fat                            & 0.050  & 5.2E-01          &                \\
KcalOutYesterday               & -0.032 & 5.3E-01          & 0.00           \\
\end{tabular}
\end{table}
\begin{table}[]
\begin{tabular}{llll}
\textbf{feature} & \textbf{corr. coeff.} & \textbf{p-value} & \textbf{reg. weight} \\
PressOutMean()Yesterday        & 0.032  & 5.3E-01          & 0.00           \\
VO2Max                         & 0.030  & 5.7E-01          & 0.00           \\
CommitsYesterday               & -0.030 & 5.7E-01          & 0.00           \\
WalkingMinYesterday            & -0.029 & 5.8E-01          & 0.00           \\
SleepRestless                  & 0.029  & 5.9E-01          & 0.00           \\
AsleepMinYesterday             & -0.027 & 6.2E-01          & 0.00           \\
SleepDeepYesterday             & 0.026  & 6.4E-01          & 0.00           \\
StepsYesterday                 & -0.025 & 6.4E-01          & 0.00           \\
Protein                        & 0.037  & 6.5E-01          &                \\
CholesterolYesterday           & 0.036  & 6.6E-01          &                \\
PressOutMax()Yesterday         & 0.022  & 6.9E-01          & 0.00           \\
SleepEfficiency                & 0.022  & 6.9E-01          & 0.00           \\
Commits                        & 0.022  & 6.9E-01          & 0.00           \\
Screentime                     & 0.022  & 6.9E-01          & 0.00           \\
SleepPointsYesterday           & 0.035  & 6.9E-01          & 0.00           \\
WindOutMax()Yesterday          & 0.021  & 7.0E-01          & 0.00           \\
SleepDurationScore             & 0.021  & 7.0E-01          & 0.00           \\
PressOutMin()                  & 0.020  & 7.1E-01          & 0.00           \\
TempWristNight                 & 0.031  & 7.3E-01          & 0.00           \\
HighLatitudeYesterday          & -0.018 & 7.3E-01          & 0.00           \\
WindOutMean()                  & 0.018  & 7.4E-01          & 0.00           \\
SleepDurationScoreYesterday    & 0.018  & 7.5E-01          & 0.00           \\
Thursday                       & 0.017  & 7.5E-01          & 0.00           \\
WednesdayYesterday             & 0.017  & 7.5E-01          & 0.00           \\
ThursdayYesterday              & 0.015  & 7.8E-01          & 0.00           \\
Friday                         & 0.015  & 7.8E-01          & 0.00           \\
WindOutMin()                   & -0.012 & 8.4E-01          & 0.00           \\
WalkingMin                     & 0.012  & 8.4E-01          & 0.00           \\
ProductiveScreentime           & 0.011  & 8.6E-01          & 0.00           \\
SleepRestlessYesterday         & 0.010  & 8.6E-01          & 0.00           \\
FloorsYesterday                & 0.010  & 8.7E-01          & 0.00           \\
TempWristNightYesterday        & 0.015  & 8.8E-01          & 0.00           \\
PressOutMean()                 & 0.009  & 8.8E-01          & 0.00           \\
MondayYesterday                & 0.008  & 8.8E-01          & 0.00           \\
Tuesday                        & 0.008  & 8.8E-01          & 0.00           \\
MoonIlluminationYesterday      & 0.007  & 9.0E-01          & 0.00           \\
MoonIllumination               & 0.006  & 9.1E-01          & 0.00           \\
Wednesday                      & -0.006 & 9.1E-01          & 0.00           \\
TuesdayYesterday               & -0.006 & 9.1E-01          & 0.00           \\
SleepEnd                       & -0.005 & 9.2E-01          & 0.00           \\
WindOutMax()                   & 0.004  & 9.5E-01          & 0.00           \\
PressOutMax()                  & 0.003  & 9.5E-01          & 0.00           \\
NapYesterday                   & 0.002  & 9.8E-01          & 0.00           \\
Nap                            & 0.000  & 9.9E-01          & 0.00           \\
ResponsivenessPointsYesterday  & 0.001  & 9.9E-01          & 0.00          
\end{tabular}
\end{table}

\end{document}